\documentclass[12pt]{article}
\usepackage{amsmath}
\usepackage{amsthm}
\usepackage{amscd}
\usepackage{amsfonts}
\usepackage[psamsfonts]{amssymb}
\usepackage{graphicx}
\usepackage{hyperref}
\def\be{\begin{equation}}
\def\ee{\end{equation}}
\def\bea{\begin{eqnarray}}
\def\eea{\end{eqnarray}}

\def\sp{\,\,\,,\,\,\,}
\def\s{\sigma}

\newcommand\beq{\begin{eqnarray}}
\newcommand\eeq{\end{eqnarray}}
\def\r{\rho}
\def\s{\sigma}
\def\fp{{\cal F}_+}
\def\fm{{\cal F}_{-}}
\def\bb{{\cal B}}

\def\jp{\langle J^+\rangle}
\def\aa{{\cal A}}
\def\hri#1#2{\href{http://arxiv.org/abs/#1}{[ArXiv:#1]#2}}
\def\hre#1#2{\href{http://arxiv.org/abs/#1/#2}{[ArXiv:#1/#2]}}

\begin{document}

\hfill{CCTP-2010-25}
\begin{center}
 \LARGE{\mbox{\bf Holographic quantum criticality and}}\\
  \LARGE{\mbox{\bf strange metal transport}}\\
 ~\\
{\large\bf Bom Soo Kim$^{a}$, Elias Kiritsis$^{b,c}$ and Christos Panagopoulos$^{b,d,e}$}\\
\mbox{\small { $^a$} Raymond and Beverly Sackler School of Physics and Astronomy,}\\ \vspace{-0.2in}
\mbox{\small   Tel Aviv University, 69978 Tel Aviv, Israel;} \\ \vspace{-0.2in}
\mbox{\small { $^b$}  Crete Center for Theoretical Physics, Department of Physics,}\\ \vspace{-0.2in}
\mbox{\small   University of Crete, 71003 Heraklion, Greece;} \\ \vspace{-0.2in}
 \mbox{\small $^c$ APC, AstroParticule et Cosmologie, Universit\'e Paris Diderot, CNRS/IN2P3,}\\ \vspace{-0.2in}
\mbox{\small  CEA/IRFU,
Observatoire de Paris, Sorbonne Paris Cit\'e, }\\ \vspace{-0.2in}
\mbox{\small   10, rue Alice Domon et L\'eonie Duquet, 75205 Paris
Cedex 13, France }\\ \vspace{-0.2in}
 \mbox{\small $^d$ IESL-FORTH, 71110 Heraklion, Greece;} \\ \vspace{-0.2in}
 \mbox{\small $^e$ Division of Physics and Applied Physics,}\\ \vspace{-0.2in}
 \mbox{\small Nanyang Technological University, 637371 Singapore}
\end{center}

\bigskip

\begin{abstract}

A holographic model of a quantum critical theory at a finite but low temperature, and finite density is studied.
The model exhibits non-relativistic z=2 Schr\"odinger symmetry and is realized by the Anti-de-Sitter-Schwarzschild
black hole in light-cone coordinates.  Our approach addresses the electrical conductivities in the presence or absence
of an applied magnetic field and contains a control parameter that can be associated to quantum tuning via
charge carrier doping or an external field in correlated electron systems.  The Ohmic resistivity, the inverse Hall angle,
the Hall coefficient and the magnetoresistance are shown to be in good agreement with experimental results
of strange metals at very low temperature.  The holographic model also predicts new scaling relations
in the presence of a magnetic field.
\end{abstract}

\newpage
\normalsize
\tableofcontents


\section{Introduction}

Strongly correlated electron systems have challenged traditional condensed matter paradigms
with weakly interacting quasiparticles \cite{Anderson:1984}.  Meanwhile, theory tools originating
from high-energy physics have been useful in addressing the physical properties of these materials,
(for a review see \cite{Sachdev:2010ch}). For example, the anti de-Sitter / Conformal Field Theory (AdS/CFT)
correspondence has proved successful in the investigation of strong-coupling
gauge theories \cite{Maldacena:1997re} with its first application focusing on conformally invariant theories.

Other non-relativistic scaling symmetries have been proposed in the context of holography involving Schr\"odinger
symmetry \cite{Son:2008ye, Balasubramanian:2008dm} or Lifshitz symmetry \cite{klm}.
The progress in geometric realizations of Schr\"odinger symmetry, with a general dynamical exponent z,
aimed for condensed matter applications has paved the way to finite temperature generalizations
\cite{Herzog:2008wg, Maldacena:2008wh, Adams:2008wt, Yamada:2008if}
using the null Melvin twist \cite{Alishahiha:2003ru, Gimon:2003xk}.

AdS space in the light-cone frame (ALCF) with z=2, has also been put forward \cite{Goldberger:2008vg, Barbon:2008bg},
as such a holographic background and a corresponding Schwarzschild black hole solution have been considered
\cite{Maldacena:2008wh, Kim:2010tf}.
Notably, while Schr\"odinger space and ALCF yield the same thermodynamic properties \cite{Herzog:2008wg, Maldacena:2008wh, Yamada:2008if, Kim:2010tf}
and transport coefficients (when the latter are independent of an embedding scalar)
\cite{Ammon:2010eq, Kim:2010tf}, ALCF is simpler and has a well-defined holographic renormalization.

Here we will analyze and report on the transport properties of ALCF, matching several universal experimental results of
the normal-state of cuprates superconductors at very low temperatures, which have been a subject of
intensive research and yet remain largely unexplained over the past two decades.
While there are other types experimental data available, such as spectroscopy and thermodynamic data,
we choose to analyze transport data because an understanding of the normal state transport properties of high 
$T_c$ cuprates is widely regarded as a key step towards the elucidation of the pairing mechanism for high-temperature
superconductivity \cite{Hussey:2008review}.

The holographic model we present provides a novel paradigm for the normal state of strange metals,
in particular high-temperature superconducting (high $T_c$) cuprates in the overdoped region,
where the charge carriers added to the parent insulator exceed the value necessary for optimal superconductivity.
Further to describing the puzzling normal state properties of these materials,
our approach leads to new falsifiable predictions for experiment.
In particular, we successfully describe the $T+T^2$ behavior of the resistivity
in \cite{Cooper2009} and the $T+T^2$ behavior of the inverse Hall angle observed in \cite{mackenzie}
at {\em very low temperatures} $T<30K$, where a single scattering rate is present.

{ This newly emerging very low temperature scaling behaviors of magnetotransport properties
are in accord with the distinct origin of the criticality at very low temperatures advertised in \cite{Hussey2009},
while the higher temperature, $T>100K$, scaling has different behaviors between
the linear temperature resistivity and the quadratic temperature inverse Hall angle, signaling two
scattering rates \cite{Tyler1997}. In searching for quantum criticality at zero temperature and
its possible connection to the origin of superconductivity, we concentrate on the lower
temperature regime with a single scattering process. We also comment on how two scattering processes emerge
by incorporating other mechanisms present in our model.
 }

In addition to the resistivity and inverse Hall angle, very good agreement is also found with experimental
results of the Hall Coefficient, magnetoresistance and K\"ohler rule on various high $T_c$ cuprates
\cite{Cooper2009,mackenzie,Hussey2009,Tyler1997,Ong1991,Takagi1992,Kendziora1992,hwang,harris,Hussey1996,tyler,Naqib2003,Nakajima2004,E1,E2,AndoBoebinger,Daou2009}.
To the best of our knowledge, no other model that describes all of these observables successfully.
Our model provides a change of paradigm from the notion of a quantum critical point,
as it is quantum critical at $T\to 0$ on the entire overdoped region.
In this sense our work breaks apart from other holographic approaches \cite{Cubrovic:2009ye, Liu:2009dm, Faulkner:2009wj},
where the measured transport is due to loop fermion effects. As such, it is applicable to a more general class of materials
{\it e.g.,} $d$ and $f$-electron systems, where the low temperature resistivity varies as $T + T^2$ \cite{stewart} and
exhibit a quantum critical line \cite{Cooper2009,zaum}.

There have been several works that use holographic approaches in order to model strange metal behavior.
The fermionic structure of such systems in the IR has been analyzed in  \cite{Cubrovic:2009ye, Liu:2009dm, Faulkner:2009wj}
and modifications due to dipole couplings in \cite{dipole}.
In particular, it was found that there is an IR scaling symmetry that could allow the realization of a marginal Fermi liquid.
The IR exponent would need, however, to be tuned for this to take place.

The linear temperature dependence of the Ohmic resistivity was realized in spaces with AdS or Lifshitz scaling
\cite{Faulkner:2010zz, Hartnoll:2009ns, cgkkm, Lee:2010ii}
and in Schr\"odinger space \cite{Ammon:2010eq,Kim:2010tf}.
A linear resistivity and a crossover to quadratic behavior was found in a larger class of scaling geometries in
\cite{cgkkm}. In the same reference, the full set of possible holographic non-trivial
low temperature behavior was classified and, as shown in \cite{qc}, comprises all possible classes of quantum critical behavior in theories with a single scalar IR relevant operator dominating the dynamics.
Finally, the temperature behavior of the Hall angle was addressed
using Lifshitz type metric with broken rotational symmetry \cite{Pal:2010sx}.

{
In section \ref{sec:SchrGeometry}, we provide the basic information for the gravity background, including how to
interpret the background compared to the extensively studied AdS.  Then,  we provide detailed properties and calculations of the
transport data using the probe DBI technique in section \ref{sec:transport}.  Magnetotransport coefficients are calculated
and analyzed in section \ref{sec:HallTransport}, where we also include the analysis of higher-temperature transport properties.
Our data is compared to the experimental results available in the literature, focusing on the universal features in section \ref{exp}.
}

\section{Holography and AdS/CFT for strongly correlated electrons}

Strong interactions of realistic finite-density systems have provided an arena for a wealth of techniques, geared
to assess in most cases the qualitative physics. A wide range of unsolved problems remain to be addressed,
especially in the realm of strange metals including condensed matter systems on the border with magnetism.
There is, therefore, an inviting opportunity for new techniques and approaches to contribute in these challenging
problems in modern condensed matter.
An interdisciplinary approach towards this aim is the utilization of the gauge-gravity correspondence,
abstracted from the correspondence between non-abelian gauge theories and string theories.
So far it has been explored in several directions, providing a novel perspective both in the modelization as well as
solution of some strongly coupled QFTs.
The hope behind potential applications to condensed matter physics is that IR strong interactions of the Kondo type in materials,
where spins can interact with electrons, may provide bound states that behave in a range of energies as non-abelian
gauge degrees of freedom that may also be coupled to other fields. The gauge interactions are characterized
by a number of charges $N_c$ that are conventionally called ``colors". Their actual number depends on the problem
at hand but it is typically small.

If this is the case, then in terms of the electrons and spins, the YM fields are composite. In the regime where the effective
YM interaction is strong, the physical degrees of freedom are expected to be colorless bound states.
Their residual interactions, analogous to nuclear forces in high-energy physics, are still strong.
{
On the other hand, the effective interaction between colorless bound states can be made arbitrarily weak
in the limit of a large number of colors, $N_c\to \infty $, as it is controlled by $1/N_c\to 0$,
although the original interaction of colored sources is strong.
}
In this limit, the theory is simplified and may be calculable.
Of course, typically, the original problem has a finite and sometimes small number of effective colors.
The question then is: how reliable are the large $N_c$ estimates for the real physics of the system?
The answer to this varies, and we know many examples in both classes of answers.
A good example on one side is the fundamental theory of strong interactions,
Quantum Chromodynamics based on the gauge group SU(3), indicating $N_c=3$ colors.
It is by now established that for many aspects of this theory, $3\simeq \infty$,  the accuracy varies in the range $3-10\%$.
It is also known that the analogous theory with two colors, SU(2), has some significant differences from its $N_c\geq 3$ counterparts.
There are other theories where the behavior at finite $N_c$ is separated from the $1/N_c$ expansion
by phase transitions making large $N_c$ techniques essentially inapplicable.

Notably, large $N_c$ techniques have been applied to strongly coupled systems for several decades,
and it is therefore natural to ask for the new contribution delivered in the present effort.
In adjoint theories in more than two dimensions, it is well known that until recently
even the leading order in $1/N_c$ could not be computed.
Although some qualitative statements could be made in this limit, the number of quantitative results was rather scarce.
On the other hand, 't Hooft observed that the leading order in $1/N_c$ is captured by the classical limit of a quantum string theory \cite{hooft}.
Finding and solving this classical string theory was therefore equivalent to calculating the leading order result in $1/N_c$ in the gauge theory.
Unfortunately, such string theories, dual to gauge theories, remained elusive until 1997, when Maldacena \cite{maldacena} made a rather
radical proposal:
(a) This string theory lives in more dimensions than the gauge theory\footnote{This unexpected (see however \cite{polyakov}) fact can be
intuitively understood in analogy with simpler adjoint theories in 0 or 1 dimensions.
There it turns out that the eigenvalues of the adjoint matrix in the relevant saddle point become continuous in the large $N_c$ limit,
and appear as an extra dimension. In general how many new dimensions may emerge in a given QFT in the large $N_c$ limit
is not a straightforward question to answer, although exceptions exist.};
(b) At strong coupling, it can be approximated by supergravity, a tractable problem.
The concrete example proposed contained on one hand a very symmetric, scale invariant, four-dimensional gauge theory
(N=4 super Yang-Mills), and on the other a ten-dimensional IIB string theory
compactified on the highly symmetric constant curvature space AdS$_5\times \mathbf S^5$.
Therefore, this correspondence becomes to be known as the AdS/CFT, or holographic, correspondence.

Although this claim is a conjecture, it has amassed sufficient evidence to spark considerable theoretical work exploring
the ramifications of the correspondence, for the dynamics on one hand of strongly coupled gauge theories and
on the other hand of strongly curved string theories.

An important evolution of the holographic correspondence is the advent of the concept of Effective Holographic Theories
(EHTs) \cite{cgkkm} in analogy with the analogous concept of Effective Field Theories (EFTs) in the context of QFT\footnote{
There are several works that contain a version or elements of the idea of the EHT \cite{rg}, although they vary in the focus or philosophy.}.
The rules more or less follow those of EFTs with some obvious differences and most importantly with less intuition.

In standard EFTs, there are several issues that are relevant:
(a) Derivation of the low energy EFT from a higher energy theory;
(b) Parametrization of the interactions of an EFT, and their ordering in terms of IR relevance;
(c) Physical Constraints that an EFT must satisfy.
Although the Wilsonian approach has allowed a good understanding of EFTs, there are still general questions
which can not be answered with our tools, for instance whether a given EFT can arise as the IR limit of a UV complete QFT.

In the context of holographically dual string theories, many issues are still not fully understood.
First and foremost is that the classical string theories dual to gauge theories cannot yet be solved.
The only approximation making these tractable is the (bulk\footnote{We refer to as the ``bulk", the spacetime in which strings propagate.
This is always a spacetime with a single boundary. The boundary is isomorphic to the space on which the dual quantum field theory
(gauge theory) lives.}) derivative expansion.
This reflects the effect of the string oscillations on the dynamics of the low-lying string modes.

It is known in many cases and widely expected that such an expansion is controlled by the strength of the QFT interactions.
In the limit of infinite strength,  the string becomes stiff and the effects of string modes may be completely neglected.
The theory then collapses to a gravitational theory coupled to a finite set of fields.
Since we are working to leading order in $1/N_c$, the treatment of this theory is purely classical. Observables (typically boundary observables
corresponding to correlators of the dual CFT)  are computed by solving second-order non-linear differential equations.

The effects of finite but large coupling are then captured by adding higher-derivative interactions in the gravitational action.
Note that this derivative expansion is not directly related to the IR expansion of the dual QFT.

The bulk theory, as mentioned earlier, has usually more dimensions compared to those of the dual QFT.
One of them is however special: it is known as the ``holographic" or ``radial" dimension, and controls the approach to the boundary of the bulk spacetime. Moreover, it can be interpreted as an ``energy" or renormalization scale in the dual QFT.

The second order equations of motion of the bulk gravitational theory, viewed  as evolution equations in the radial direction,
can be thought of as Wilsonian RG evolution equations \cite{deboer}. The boundary of the bulk spacetime corresponds to
the UV limit of the QFT. Although the equations are second order they need only one boundary condition in order to be solved,
as the second condition is supplied by the ``regularity" requirement of the solution at the interior of spacetime.
Here gravitational physics proves particularly helpful: a gravitational evolution equation with arbitrary boundary data
leads to a singularity. Demanding regular solutions gives a unique or a small number of options.
The notion of ``regularity" can however vary, and may include runaway behavior as in the case of holographic
open string tachyon condensation relevant for chiral symmetry breaking \cite{chi}.

The holographic model and associated saddle point we will explore here is rather simple and does not require a very sophisticated machinery.
It has, however, a non-relativistic Schr\"ondiger symmetry, and this is a realm that has not been explored fully so far.

\section{ Schr\"odinger geometry} \label{sec:SchrGeometry}

The model we present is comprised of two sectors. The first is gravitational and contains the metric as a single field. It controls the dynamics of energy in the theory, and we will analyze it in this section. The second contains the dynamics of the charge carriers and will be given by a Dirac-Born-Infeld (DBI) action of a
 gauge field dual to the conserved current of the carriers. We will analyze this part in a later section where we will calculate the conductivities.

The gravitational action is the Einstein action with a negative cosmological constant
\begin{align}\label{eq:OriginalAct}
  I = \frac{1}{16\pi G_5} \int d^5x\sqrt{-g}
  \bigg( \mathcal{R} + \frac{12}{\ell^2} \bigg) \;,
\end{align}
where the symbols $g$, $\mathcal{R}$ and $\ell$ are the determinant of the metric,
the scalar curvature and the length scale of the theory related to
the cosmological constant, respectively. We suppress the boundary terms needed for
proper boundary conditions and renormalization, and consider the AdS-Schwartzschild
black hole solution in light-cone coordinates
 \cite{Maldacena:2008wh, Kim:2010tf}
\begin{align}
  ds^2 =& g_{++} dx^{+2} + 2 g_{+-} dx^+ dx^- + g_{--} dx^{-2} + g_{yy} d y^2 + g_{zz} d z^2 + g_{rr} dr^2 \;,
  \label{AdSinlightcone}
\end{align}
where
\begin{gather}
  g_{++} =  \frac{(1-h) r^2}{4b^2 \ell^2} \,, \quad
  g_{+-} = - \frac{(1+h)r^2}{2 \ell^2}  \,,  \quad  g_{--} =  \frac{(1-h)b^2 r^2}{\ell^2} \,,\; \quad
  g_{yy}  =  g_{zz} =  \frac{r^2}{\ell^2} \,, \nonumber\\
  g_{rr} =  \frac{\ell^2}{h r^2}  \,, \quad
    h = 1 - \frac{r_H^4}{r^4}\;, \quad x^+ =b(t+x)   \;,\quad
  x^- = \frac{1}{2b}(t-x)
  \;.
  \label{MetricComponents}
\end{gather}
To ensure z=2, we assign $[b]$ (the scaling dimension of $b$ in the unit of mass) as $-1$, and thus
$[x^+] = -2$ and $[x^-]= 0$.
The full 10-dimensional space, AdS$_5 \times S^5$,  in light-cone coordinate was written in {\it e.g.,}
\cite{Ammon:2010eq}\cite{Kim:2010tf}.
We drop the $S^5$ part for the rest of our discussion,
except for the embedding scalar discussed below,
because it is decoupled and becomes an overall factor
in the probe brane DBI action \cite{footnote1}.

To match the non-relativistic isometry group, one of the light-cone directions, $x^+$ with scaling dimension $-2$,
is identified as time, and we fix the momentum of the other light-cone coordinate, $x^-$ \cite{Son:2008ye, Barbon:2008bg}.
The thermodynamic properties of the ALCF are identical to those of Schr\"odinger space \cite{Herzog:2008wg, Maldacena:2008wh,
Yamada:2008if, Kim:2010tf}, explained below in section \ref{sec:roleofb}. The interpretation of this coordinate system
is connected to being on the infinite momentum frame along a single spatial direction, which we take here as $x$.
In this frame, the nontrivial physics occurs in the two transverse spatial dimensions $y,z$.

\subsection{Schr\"odinger geometry and its interpretation}

The initial geometry is the AdS Schwartzschild black hole, which is known to describe a strongly coupled Conformal Field Theory (CFT) at finite temperature.
However, here it is described in the light-cone coordinate system and since $x^{+}$  will be taken as time, the symmetry is
broken to a z=2 Schr\"odinger symmetry. In this sense, the bulk background, equations (\ref{eq:OriginalAct})-(\ref{MetricComponents}),
describe the strongly coupled "glue" that interpolates between conformal symmetry at high temperatures and z=2 Lifshitz like
non-relativistic scaling symmetry near $T=0$.

A qualitative way to understand this is to appreciate that in these coordinates the ``speed of propagation" of signals
in the bulk spacetime asymptotes to zero as we approach the black-hole horizon.
This is a well known effect in black hole space-times \cite{k} in this coordinate system - also known as the Carolean limit.

This transition, from AdS critical (z=1) to Lifshitz critical (z=2), is a key ingredient of the gravitational black hole background.
It is important to identify where the transition occurs.  In the bulk background, this is controlled by the parameter $b$. {Here}, $b$ is a length scale that parametrized precisely this transition,
in a way that preserves scale covariance.
{In brief}, the bulk geometry is an interpolation between (z=1) and (z=2) geometries in the IR.
The associated dual theory should likewise interpolate between two energy regimes, one where it has the usual relativistic scale symmetry and another where it has the Lifshitz symmetry.

It should be noted that the gravitational background is 5 dimensional. Apart from the holographic directions, there is a time direction and 3 regular space directions $x,y,z$.
In light cone coordinates, $x^{\pm}=x\pm t$, one of the spatial coordinates, {namely} $x$, is playing a special role.
The Schr\"odinger frame can be considered as an infinite boost in the $x$ directions (this is the infinite momentum frame in QFT)
as we {discuss} below. In this limit, all dependence on $x$ spacial direction is redundant, hence {the} physics depends only on two spatial directions $y,z$.
Therefore, it is these two spacial directions that the theory {depends upon}, and the dual quantum field theory is 2+1 dimensional.

\subsection{The role and interpretation of the parameter $b$}\label{sec:roleofb}

There are two control parameters in this model, $b$ and $E_b$, which will be introduced in the following section.
Both are dimensionful but can form dimensionless combinations either alone or combined with temperature.

The significance of the parameter $b$ can be appreciated physically from the thermodynamics of
the same system described in \cite{Maldacena:2008wh}\cite{Kim:2010tf}. These are as follows:
\begin{align}
  E = \frac{\pi^3 \ell^3 b^4 T^4 V_3}{16 G_5 }
  \;,\quad
  J = - \frac{\pi^3 \ell^3 b^6 T^4 V_3}{4 G_5}
  \;,\quad
  S = \frac{\pi^3 \ell^3 b^4 T^3  V_3}{4G_5}
  \;, \quad
  \Omega_H = \frac{1}{2b^2}
  \;,
\label{sc}\end{align}
where we have defined $V_3 := \int dx^-dydz$ and used $r_H = \pi \ell^2 b T$.
$\ell$ is the AdS length, while $G_5$ is the five-dimensional Newton's constant. $J$ is the charge associated with the translational symmetry in $x^-$, that is conserved in the Schr\"odinger geometry, while   $\Omega_H$
is the associated chemical potential.

To understand the non-relativistic z=2 scaling, the mass dimensions of various parameters are  $[b]=-1$, $[x^+] =-2$,
$[x^-]=0$, $[y]=[z]=-1$ and $[V_3] =-2$.
From $[G_5] =-3$, we obtain $[J]=0$, $[\Omega_H] =2$, $[M]=2$, $[S]=0$, $[\beta]= -2$ and $[T]=2$.
These are consistent with the dimensions of the non-relativistic systems with the dynamical exponent z=2, as described in appendix F in \cite{Maldacena:2008wh}.

Therefore, the parameter  $b$ can be associated with the chemical potential for the conserved particle number of the Schr\"odinger symmetry.
The dimensionless quantity $b^2T$ is associated with the crossover behavior between the z=1 and z=2 regimes of the black hole solutions.

It can be seen from (\ref{sc}) that $b$ controls also the system's response to external pressure. Therefore, different values of $b$ correspond to different external pressures for the ``glue" ensemble. External pressure is a widely used quantum tuning parameter to study the evolution of the ground state electronic properties in a range of strange metals including organic superconductors, heavy fermion systems and other strongly correlated electron systems.

\section{Holographic DBI transport}\label{sec:transport}

We will now add to the system, charge carriers, using D-branes.
To calculate the transport properties, we follow the standard DBI probe approach
\cite{Karch:2007pd}\cite{Ammon:2010eq}\cite{Kim:2010tf}.
We introduce $N_f$ D7 branes on the background and work in the probe limit,
$N_f\ll N_c$.
The D7 branes cover three angular directions $S^3$ of $S^5$ in addition to the background (eq. (\ref{AdSinlightcone})).
From this embedding there are two remaining world volume scalars on the branes. One scalar is chosen to be
trivially constant and the other a function of the radial coordinate $\theta (r)$. Hence, D7 has
the same metric as eq. (\ref{AdSinlightcone}) with a simple modification $g_{rr} \rightarrow
g_{rr}^{D7} =  g_{rr} + \theta' (r)^2$.

 We consider the $U(1)$ world-volume gauge field $A_{\mu}$, which is dual to the conserved current $J^{\mu}$ of the charge carriers.
To have an electric field $E_b$ only along the $x^+$ direction, we choose the gauge fields as
\begin{gather}
	A_+ = {E_b\over 2 \pi \ell_s^2 } y + h_+ (r) \;, \quad A_- = {b^2 E_b\over  \pi \ell_s^2} y + h_- (r) \;, \quad
	A_y =  {E_b b ^2\over \pi \ell_s^2} x^- +  {h_y} (r) \;.
\end{gather}
The light-cone electric field is a vector. We turn it on in one direction only (the $y$ direction above).
The system, however, is rotationally invariant despite appearances for reasons that are explained
 in appendix \ref{sec:rotationInvariance}, along with more detailed calculation of the transport.
The resulting probe DBI action has the form
\begin{equation}
	S_{D7} = - N_f T_{D7} \int d^8 \xi  \sqrt{-\det (g_{D7} + 2\pi \ell_s^2 F)} \ ,	\label{ProbeAction}
\end{equation}
where $T_{D7}, \xi$ and $F$ are the D-brane's tension,
the world-volume coordinate and the $U(1)$ field strength, respectively.

There are three constants of motion, which we identify as three currents
$\langle J^\mu \rangle = \frac{\delta {\cal L}}{\delta h_\mu'}$, where $\mu = +, -$ and $y$.
We solve the equations of motion in terms of these currents, and obtain the on-shell action along the lines of
\cite{Karch:2007pd}.  Furthermore, we demand the square root in the action to be real all the way from the horizon, located at $r=r_H$,
to the boundary at infinity. As shown in appendix \ref{sec:rotationInvariance}, it delivers two important relations:
\be
\langle J^- \rangle = -\frac{g_{+-}(r_*) }{g_{--}(r_*) } \langle J^+ \rangle\;,
\ee
and Ohm's law, $\langle J^y \rangle = \sigma E_b$, with
\begin{align}
	&\sigma=\sigma_0 \sqrt{{J^2\over t^2 A(t)}+{t^3\over \sqrt{A(t)}}}\;, \qquad \; A(t)=t^2+\sqrt{1+t^4} \;.
	\label{conductivity}
\end{align}
where $\sigma_0={ {\cal N} b\cos^3 \theta\sqrt{2b E_b}}$, and we use the dimensionless scaling variables
\begin{equation}
t={\pi\ell Tb\over \sqrt{2b  E_b}}\;\;,\;\;J^2={64\sqrt{2} \langle J^+ \rangle ^2\over
(  {\cal N} b\cos^3 \theta)^2 (2b E_b)^3}.
\label{cond}\end{equation}

Equation (\ref{conductivity}) is particularly interesting in the regime
$t\ll 1$, $J\gg 1$ ;
\begin{equation}
\rho=\frac{1}{\sigma} \approx {t\over J\sigma_0}=
\frac{ \pi \ell b \sqrt{E_b b }}{ \langle  J^+ \rangle }~ T \;.
\end{equation}
Therefore, the Ohmic resistivity is linear in temperature in the  low-$T$ regime of the model.

\begin{figure}[!ht]
\begin{center}
	 \includegraphics[width=0.55\textwidth]{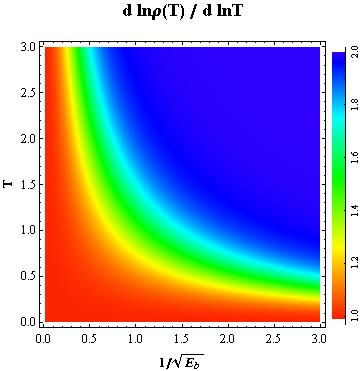}
	 \caption{The exponent of $\frac{d \ln \rho (T)}{d \ln T}$
	 as a function of a tuning parameter $\frac{1}{\sqrt{E_b}}$ and temperature $T$ at low temperatures.
	 Note that the linear temperature dependence of the resistivity extends over the low temperature range, with
	 $\rho \sim T + T^2$.
	 Compare this plot to Fig. 3 of \cite{Cooper2009}.
	 }
	 \label{fig:LCResT}
\end{center}
\end{figure}

We now focus on the first term of eq. (\ref{conductivity}).
At low temperatures, this term dominates over the second one, namely when $t\ll J^{1\over 3}$, $J\gg 1$.
Notably, the first term is due to the drag force exerted by the medium on heavier charge carriers (drag limit)
\cite{Karch:2007pd}.
In this limit, the resistivity reads
\begin{align}
	\rho \approx {t\over J\sigma_0}\sqrt{t^2+\sqrt{1+t^4}} \;.
	\label{DragLimit}
\end{align}
The drag mechanism here is purely stringy and is explained  below in subsection \ref{sec:SCTM}.

By increasing the scaling variable $t$, the temperature
 dependence of the resistivity crosses from linear
  $\rho \approx {t\over J\sigma_0}$ to quadratic $\rho \approx \frac{\sqrt{2}~t^2}{J\sigma_0}$.
This crossover is governed by the bulk parameter $b$, setting the scale of the Lifshitz symmetry.
$E_b$, on the other hand, is a more interesting parameter. Its direct physical interpretation is not
straightforward  as it is the light-cone component of an electric field in the boost direction $x$.
{ In section \ref{sec:RoleOfEb}, }we also explain the interpretation of
 $E_b$  and discuss why we expect   $E_b\to 0$ to correspond to the heavily overdoped region whereas $E_b\to\infty$ to  optimal doping.
The crossover behavior observed is due to the fact that effectively the gravitational background  (\ref{AdSinlightcone}),
interpolates between z=1 (AdS) symmetry in the UV to z=2 Lifshitz symmetry in the IR.

\subsection{Strong-coupling transport mechanisms}\label{sec:SCTM}

We will now comment on the resistivity results discussed in the previous section.

In strongly coupled systems described holographically, the conductivity of charge carriers has typically two contributions
(that add quadratically), the ``drag" term and the ``pair-creation" term \cite{Karch:2007pd}.

The physical picture corresponding to the drag contribution is that a charged ``quark"
moves through the strongly coupled (glue) plasma dragging behind its flux that is represented
in the (fundamental) string. The ``string" should be considered as the glue field attached to the ``quark".
There is a world-volume horizon on that string, which has been interpreted to separate the part of the
tail that has thermalized via interactions with the plasma and that is closer and follows the ``quark".
 It is loosing energy because of the strong interactions with the plasma.

A Drude-like formula relates this energy loss and terminal velocity to the conductivity (drag conductivity).
Although the Drude formula is classical and its physics are well understood, the result of the energy loss at strong
coupling is poorly understood.
The same mechanism for QCD is more or less experimentally tested in heavy-ion collisions. However,
there is no alternative theoretical understanding of the dependence of the energy loss on terminal velocity, etc.,
apart from general symmetry considerations.
The clear picture exists in the gravitational description: the resistance is due to the energy
loss of a string moving in the appropriate gravitational background.

The other contribution is {expected to be due to} light charged pairs created from the vacuum contributing to the conductivity.
This contribution is Boltzmann suppressed and controlled, in our model (and in \cite{Karch:2007pd}), by the coefficient
${\cal N}$ given in equations (\ref{conductivity}) and (\ref{cond}) above.
In full blown holographic models, this depends explicitly on the
UV mass of charge carriers. Notably, this contribution comes from strong
coupling and no alternative calculations of this exist in the same regime for comparison.
This term picks up at higher temperatures and is not relevant to the regimes discussed {below}.
Here, we are interested in the very low temperature regime to study the
possible presence of quantum criticality and the associated
superconducting mechanism.

Therefore, the model includes a bulk geometry representing critical ``glue" that crosses over from z=1 to z=2
behavior in the IR and  massive charge carriers (as probes) moving in this background, losing energy via
the ``drag" strong coupling mechanism.

\subsection{The role and interpretation of the parameter $E_b$}\label{sec:RoleOfEb}

The parameter $E_b$ controls the physics of charge transport in analogy to experimental tuning parameters
such as charge carrier doping, pressure, electric field  or in-plane magnetic field.

A priori, $E_b$ is a light-cone electric field component, $E_b=F_{{+y}}$.
More precisely, as detailed in appendix \ref{sec:rotationInvariance}, it is a vector with two components,
$E^y_b=F_{{+y}}$ and $E^z_b=F_{{+z}}$.
However, as shown there,  we may set $E_b = \sqrt{(E_b^y)^2+(E^z_b)^2}$
and describe the transport properties in terms of $E_b$ without loss of generality.

In the same appendix we also show that, despite the fact that the vector light-cone electric field is non-zero,
transport is in fact rotationally invariant.

Since $E_b$ is the only non-zero electric field component and, in particular, does not break rotational invariance
in the transverse $y,z$ directions, its presence {demands} an interpretation. Such an electric field can be obtained
by an infinite boost along the $x$ directions from a standard electric field $E_y$ in the $y$ direction.
Under a boost $\lambda=\tanh{v\over c}$ along the $x$ direction,
$$
F_{+y}'={\lambda\over 2}E_y\sp  F_{-y}'={1\over 2\lambda}E_y \;.
$$
Therefore, to arrive at our set-up we need to send $\lambda\to \infty$ and $E_y\to 0$ so that the product is finite
$$
E_b=\lim_{\lambda\to\infty\atop E_y\to 0} {\lambda\over 2}E_y
$$
Therefore, a non-zero $E_b$ reflects an infinite boost of the system in the $x$ direction and an infinitesimal
 electric field in the $y$ direction. This limiting procedure explains why we should not expect rotational invariance in the $y$-$z$ plane
 to be  broken as demonstrated explicitly in appendix \ref{sec:rotationInvariance}.

To interpret the effect of varying $E_b$, we will have to follow it through the passage to the infinite momentum frame.
This translates into varying the "speed of light" $c$ that enters in the boost.
Therefore, fixing the same infinitesimal $E_y$ in the rest frame and varying the "speed of light" is equivalent to varying $E_b$
in the infinite momentum frame - in particular as $c\to 0$, $E_b\to\infty$.
In our metric, this variation is implemented by varying the IR scale $b$ that controls the passage between z=1 and z=2 scaling
in the bulk geometry. This is also visible in all our expressions for the conductivity, in terms of the scaling variables
where $E_b$ appears always in the combination $bE_b$.
Therefore, $E_b$ should not be thought as an external field but as an internal variable parameter of the system.

By the relativity principle, we conclude that the infinite momentum frame captures the physics of charge carriers in two regimes: \\
(a) The $z=1$ CFT regime when $t={\pi\ell Tb\over \sqrt{2b  E_b}}\gg 1$. \\
(b)  The $z=2$ Liftshitz-like regime when  $t={\pi\ell Tb\over \sqrt{2b  E_b}}\ll 1$.

The transition temperature is controlled by $E_b$. $E_b\to 0$ maps to the large "doping" region where {the} resistivity is quadratic
at all scales. This is the quadratic resistivity of CFT and is not necessarily associated, as is now well known, to fermions or bosons
(in the $N=4$ example, it is both.) $E_b\to \infty$ maps to optimal doping {where} the resistivity is linear at all scales.

There are several side arguments that support this map.

1. In parametrizing the resistivity as $\rho=a_1 T+a_2 T^2$ at low temperature, experiments indicate $a_2$
to be constant and $a_1$ to decrease rapidly with doping \cite{Cooper2009}.
In our model, $a_2$ is indeed  independent of  $E_b$, while $a_1\sim \sqrt{E_b}$ and vanishes across the
"overdoped regime" ($E_b\to 0$).

2. The scaling variable for the magnetic field is ${\cal B}\sim {B_b \over E_b}$
 and the conductivities depend on ${\cal B}$ alone.
This is in accordance with experimental observations, where as one moves to the
 overdoped region the effects of the magnetic field are stronger \cite{tyler}.
  This is discussed in more detail {below.}  Notably, in the families of
  strange metals one may vary the chemical potential also using an external
   magnetic and electric field and not necessarily chemical doping.

It is not entirely clear at the moment how parameters such as the "internal light velocity"
 (as defined by the holographic metric) is related to standard physical properties
  of the material - charge density, velocity of quasiparticles, etc. To assert this,
   a more detailed analysis is necessary where several new constituents should be
   considered - for instance, the calculation of correlation functions of currents,
   couplings to fermions, and potentially others.  This analysis maybe necessary to
    provide further features for this class of ideas. However, it is beyond the
    focus of the present effort.

\section{Holographic Hall transport} \label{sec:HallTransport}

{ In this section}, we analyze the charge transport  in the presence of a magnetic field following \cite{O'Bannon:2007in}.
The detailed calculation is carried out in appendix \ref{sec:HallCal}. The analysis of the behavior of the conductivity in different regimes can be found
in appendix \ref{apb}.

The gauge fields now are
\begin{align}
	A_+ = {E_b\over 2 \pi \ell_s^2 } y + h_+ (r) \;, ~~ A_- = {b^2 E_b\over  \pi \ell_s^2} y + h_- (r) \;, ~~
	A_y =  {E_b b ^2\over \pi \ell_s^2} x^- +  {h_y} (r) \;, ~~ \; A_z = {B_b y\over  2 \pi \ell_s^2} +  h_z (r) \;.
	\label{GaugeHall}
\end{align}
This configuration includes a light-cone electric field, $E_b$,
along the $y$ direction and a magnetic field, $B_b$, perpendicular to the $y, z$ directions.
The DBI probe action eq. (\ref{ProbeAction}) has four conserved currents,
$\langle J^\mu \rangle$, related to the variation of $h_\mu' (r)$
with $\mu = +, -, y$ and $z$.
The exact Ohmic conductivity in the presence of a magnetic field is
\begin{align}
	\sigma^{yy}
	= &\sigma_0\frac{\sqrt{{\cal F}_+ J^2
	+ t^4\sqrt{{\cal F}_+} {\cal F}_- }} {{\cal F}_-} \;, \quad \;\sigma^{yz}
	=\bar\sigma_0\frac{{\cal B}}{{\cal F}_-} \;,
	 \label{electricConductivitywithB}
\end{align}
where $\bar\sigma_0=\frac{\langle  J^+ \rangle }{b E_b}$,  $\sigma_0$ was defined earlier (eq. (\ref{conductivity})), and $t, J$ in eq. (\ref{cond}).
Here
\begin{equation}
{\cal F}_\pm=  \sqrt{\left({\cal B}^2 + t^4 \right)^2
+  t^4 }  \mp {\cal  B}^2 + t^4\;, \quad \;\;{\cal B}={B_b\over 2b E_b} \;.
\label{bcond}\end{equation}
Note that eqs. (\ref{electricConductivitywithB}) and (\ref{bcond}) reduce to eq. (\ref{conductivity}) for ${\cal B}=0$.

\begin{figure}[t]
\begin{center}
\begin{tabular}{ccc}
	 \includegraphics[width=0.3\textwidth]{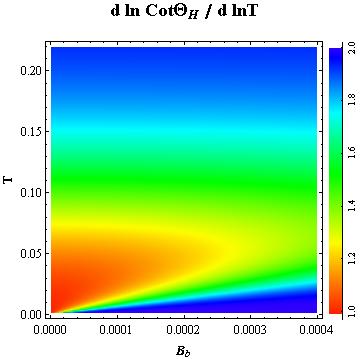} ~~
	 \includegraphics[width=0.3\textwidth]{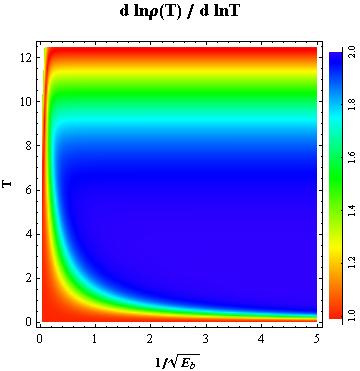} ~~
	 \includegraphics[width=0.3\textwidth]{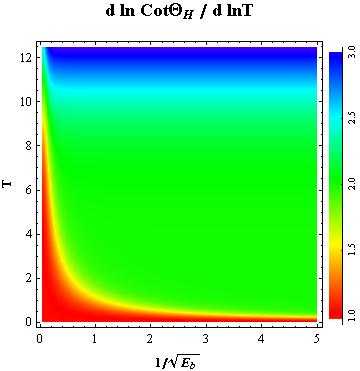}
\end{tabular}
\caption{Left: Temperature ($T$) and magnetic field ($B$) dependence of the exponent of $\cot\Theta_H$ in the low $T$, low $B$ regions. Middle: The effective power $n$ of the resistivity $\rho \sim T^n$ at zero magnetic field as a function of temperature ($T$) and the effective doping parameter $1/\sqrt{E_b}$. For $T\lessapprox 8$, the resistivity is dominated by the drag mechanism, while at $T \gtrapprox 8$ it is dominated by the pair-creation term.  Right: the effective power dependence of $\cot\Theta_H$ at small magnetic field, as a function of temperature and  $1/\sqrt{E_b}$. For $T\lessapprox 8$, the resistivity is dominated by the drag mechanism, while at $T \gtrapprox 8$ it is dominated by the pair-creation term. Note that here the range of the power varies from 1 to 3.  }
\label{BT}
\end{center}
\end{figure}

For a rotationally symmetric system with a plane of $y, z$ coordinates, the resistivity matrix is defined as the inverse of the conductivity matrix.
The inverse Hall angle is defined as the ratio between Ohmic conductivity and the Hall conductivity as
$ \cot \Theta_H =\frac{\sigma^{yy}}{\sigma^{yz}} $. We also define the Hall coefficient $R_H$ and the magnetoresistance ${\Delta\rho\over \rho}$ as
\begin{align}
R_H={\rho_{yz}\over B} \;, \quad \;
{\Delta\rho\over \rho}={\rho_{yy}(B)-\rho_{yy}(0)\over \rho_{yy}(0)} \;.
\end{align}
For $J$ sufficiently large, the resistivities are given by  drag contributions. There are three relevant regimes:\\
(a) ${\cal B}\ll t\ll 1$ with
\begin{equation}
R_H\simeq {\bar\s_0 \over \s_0^2 J^2 }\;, \qquad \;
\cot\Theta _H\simeq {\s_0 J\over \bar\s_{0} \bb}t
\;, \qquad \; {\Delta \rho\over \rho}\simeq {3\over 2}{\bb^2\over t^2} \;,
\label{InverseAngle1}\end{equation}
(b) $\bb\ll t^2$ and $t\gg 1$ with
\begin{equation}
R_H\simeq {\bar\s_0 \over \s_0^2 J^2 }\;,\qquad\;
\cot\Theta _H\simeq {\sqrt{2}\s_0 J\over \bar\s_{0} \bb}t^2\;, \qquad  \;
{\Delta \rho\over \rho}\simeq {\bb^2\over t^4} \;,
\label{InverseAngle3}\end{equation}
(c) $\bb\gg t$ and $\bb\gg t^2$ with
\begin{equation}
R_H\simeq {2\over \bar\s_0 }\;, \qquad
\cot\Theta _H\simeq {\s_0 J\sqrt{1+4\bb^2}\over \bar\s_{0} \sqrt{2}\bb^2}~t^2\;, \qquad
{\Delta \rho\over \rho}\simeq  {2\sqrt{2}\s_0^2 J^2 t^2 \over \bar\s_0^2 t {\cal A} } \;.
\label{InverseAngle2}\end{equation}
For a summary of these properties we refer the reader to Fig.  \ref{BT}.

The above-mentioned transport properties can be compared successfully to those of strange metals as described in the section below.

\section{Comparison to experiment\label{exp}}

Since the discovery of the high $T_c$ cuprate superconductors 25 years ago,  there have been significant experimental efforts to identify the physical mechanism governing their unconventional superconducting and normal state properties.  Magnetotransport has been at the heart of studying the emerging properties of superconductors.  Here, we focus our discussion on characteristic quantities which have puzzled the condensed matter community and remain largely unexplained.  We discuss especially the region where the concentration of charge carrier doping is sufficiently high to span the phase diagram from optimal superconductivity (optimal doping) towards its suppression due to excessive carrier concentration (overdoping). For this we chose to address two prototypical copper oxide superconductors, namely $La_{2-x}Sr_xCuO_4$ (LSCO) and  $Tl_2Ba_2CuO_{6+\delta}$ (TBCO), for which it is possible to span the abovementioned doping range.  The normal state of these superconductors may be accessed by suppressing superconductivity, for example, through the substitution of $Zn$ for $Cu$ (see {\it e.g.}, \cite{Ong1991,Takagi1992,Naqib2003}) or by a sufficiently high applied magnetic field  \cite{Daou2009, Cooper2009}. For a nice review concerning anomalous transport properties of cuprates, see {\it e.g.,} \cite{Hussey:2008review}.

It has been generally accepted that at optimal doping i.e., where the absolute value of the superfluid density is highest, the resistivity in the normal state of cuprate superconductors varies linearly with $T$. This unconventional behavior has often been discussed in terms of quantum criticality. As charge carrier doping increases, the linearity gives way to higher power laws and eventually a more or less Fermi liquid regime emerges. However, recent low temperature transport data \cite{Cooper2009} on LSCO have challenged earlier works \cite{boebinger2009}.  In \cite{Cooper2009},  the authors reported that the suppressed superconducting region is replaced by a "2D strange metal", with the Ohmic resistivity at low temperature behaving as $\rho \sim T + T^2$. Especially the doping region where the resistivity varies linearly with $T$ is broader than expected and continues to survive in the heavily overdoped side of the phase diagram. This result suggests a line of critical points and therefore a significant departure from our earlier understanding on the possible role of the above mentioned linearity and a well defined, singular quantum critical point coinciding with optimal superconductivity.
This result is in fact consistent with an earlier observation on TBCO at very low temperatures $T<30K$ - see Figs 5 and 6 in \cite{mackenzie}. Notably, a line of critical points has recently been argued for another group of unconventional superconductors on the border of magnetism, namely the $f$-electron systems \cite{zaum}.

The inverse Hall angle has been shown to vary as $\cot \Theta_H \sim T+ T^2$ at very low temperatures in TBCO \cite{mackenzie}, which is surprising on the basis of the conventional wisdom considering two scattering rates in the cuprate superconductors. In particular, the inverse Hall angle and the resistivity behave in a similar manner, namely as $\cot \Theta_H  \sim \rho \sim T + T^2$ at very low temperature, $T<30$.  This is clearly depicted in Fig. 9 of \cite{mackenzie}.  It has been argued that two scattering rates observed in the overdoped region of TBCO collapse on to a single scattering rate as $T \rightarrow 0$, in the temperature range $T<30K$ \cite{mackenzie}.
The similar behavior between resistivity and inverse Hall angle might be considered to be the realm of a Fermi liquid, yet their strong linear temperature dependence over a broad range of doping is a challenge
\cite{mackenzie}\cite{Cooper2009}\cite{boebinger2009}.
For LSCO, similar behaviors were observed for the inverse Hall angle \cite{hwang}.

Here, we compare the results of our model with the experimental results. We focus our attention on several key and outstanding features of the normal state of cuprate superconductors. Namely, the analysis of the in-plane resistivity, in-plane Hall coefficient, inverse Hall angle,
in-plane magnetoresistance and the modified K\"ohler rule. We start by summarizing the main features of the transport properties described by our model.

\begin{enumerate}

\item In the absence of an applied magnetic field, there is a linear resistivity near $T=0$, which changes to quadratic at higher temperatures.
The coefficient of the quadratic term is independent of $E_b$, whereas that of the linear term is proportional to $\sqrt{E_b}$,
which is directly related to the inverse of the doping.

\item In the presence of a magnetic field,   $\cot\Theta_H$  is linear when the resistivity is linear, and quadratic when the resistivity is quadratic.
This is the behavior seen in strange metals at very low temperatures (for example below 25 K in overdoped $Tl_2Ba_2CuO_{6+\delta}$).
At higher temperatures however, the quadratic behavior in real materials dominates the overdoped side.

\item The magnetoresistance calculated is in agreement with experimental data at low temperatures. The model predicts that near $T=0$ the magnetoresistance dives sharply towards zero.

\item The universal scaling behavior of Hall coefficient, available in the experimental literature, is qualitatively very similar to the scaling function
$\frac{1}{t \sqrt{A}}$ of our model.

\item The ``modified K\"ohler" rule is known to be valid for cuprates and other related materials. K\"ohler's rule has been  {shown} experimentally to fail at relatively
high temperatures in the overdoped region. It has been argued that this is due to a superconducting instability \cite{kimura}. We show that it is also compatible with the correlation between $\cot\Theta_H$ and resistivity
 {as observed} experimentally at low temperatures. The model therefore predicts  that at sufficiently low temperatures  both the K\"ohler and modified K\"ohler rules are valid in the overdoped region.

\end{enumerate}

\subsection{Resistivity}

Let us concentrate on the recent experimental observations on LSCO and TBCO at very low temperature. In overdoped TBCO the resistivity in the millikelvin regime follows $\rho\sim T+T^2$ \cite{mackenzie}. Recently, very low temperature resistivity data on LSCO over a wide range of doping, namely from slight underdoped $p=0.15$ to heavily overdoped $p=0.33$, indicate that the suppressed superconducting region in the overdoped regime has an unexpected $\rho=a_0+a_1T+a_2T^2$ behavior, with a particularly interesting linear temperature dependence of the resistivity at very low temperatures \cite{Cooper2009}. Furthermore, $a_2$ was found to be doping independent, while $a_1$ decreased rapidly with overdoping. Earlier works in the overdoped region (above $p\sim 0.2$)  for LSCO reported a novel power-law $\rho = \rho_0 + A T^n$ with $n\sim 1.5$  dominating the resistivity over a wide temperature range (see for example,  Fig. 1 in \cite{Takagi1992}). Here we make a comparison of our results to the abovementioned reports.

 We focus on the drag limit mentioned above. The drag term, proportional to $J^2$ in eq. (6), dominates in the low temperature limit. Here, the resistivity has two different contributions, one linear in $T$ and another $T^2$,
\begin{align}
	\rho \approx  a_1 T  =  \left( \frac{ \sqrt{ E_b / b}}{\ell \pi } \right)
	\frac{\ell^2 \pi^2 b^2}{ \langle   J^+ \rangle  } T  \;, \qquad
	\rho \approx   a_2 T^2 =  \frac{\ell^2 \pi^2 b^2 }{  \langle  J^+ \rangle  } T^2  \;.
\end{align}
$a_2$ is doping independent whereas $a_1$ decreases rapidly with doping, in agreement with our model.  We may therefore, map $\frac{1}{\sqrt{E_b}}$ to the doping parameter as depicted above in Fig. \ref{fig:LCResT}.

\subsection{Inverse Hall angle}

\begin{figure}[!ht]
\begin{center}
	 \includegraphics[width=0.65\textwidth]{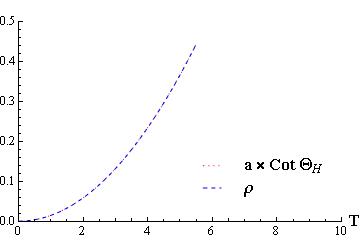}
	 \caption{Plot of the resistivity and inverse Hall angle,  in our model,  for the low-temperature regime
	 with small magnetic field. Note that
	 the inverse Hall angle has been scaled by a constant factor
	 $a= B_b / (32 \sqrt{2} \langle J^+ \rangle$. This plot is to be compared with  Fig. 9
	 of \cite{mackenzie}.
	 }
	 \label{fig:ResHallAng}
\end{center}
\end{figure}

The inverse Hall angle is defined as  $\cot\Theta _H=\s_{yy} /  \s_{yz}$.  At optimal doping and relatively high temperature ($T \geq 100 K$ for YBCO \cite{Ong1991}, LSCO \cite{hwang} and TBCO \cite{Tyler1997}),
$\cot\Theta _H$ varies universally as $T^2$, while the corresponding Hall coefficient is highly irregular.  To the best of our knowledge there is no corresponding systematic data available for optimal doping at very low temperatures.

The first observation of $\cot\Theta_H = T + T^2$ in overdoped samples at low temperatures is depicted in Fig. 8 of \cite{mackenzie}. Notably, the resistivity and the inverse Hall angle for TBCO behave in a similar manner at low temperature (Fig. 9 of \cite{mackenzie}). There is also indirect evidence for universality from works on LSCO see {\it e.g.,} Fig. 3 of \cite{Ando1997} and Fig. 3 (c) of \cite{Ando2004PRL92}. Further support may be obtained from earlier studies on overdoped LSCO.  For instance, in \cite{hwang} the authors suggest $\cot\Theta_H$ cannot be fitted by $A+BT^2$ in the range $T= 4 K$ to $T= 500 K$ (Fig. 4 in \cite{hwang}).  A thorough investigation at very low temperature however, has yet to be performed.

Our results demonstrate that the resistivity and the inverse Hall angle behave in a similar manner when the system is at low temperature and small magnetic fields, indicating that we are working in a linear regime or a weak field regime as defined by the experimental results for the magnetoresistance
$ {\Delta \rho\over \rho} \sim B_b^2$  and Hall coefficient
$R_H \sim B_b^0 \sim const.$ \cite{tyler}. This is depicted in Fig. \ref{fig:ResHallAng}.

\subsection{Magnetoresistance}

\begin{figure}[!htb]
\begin{center}
	 \includegraphics[width=0.65\textwidth]{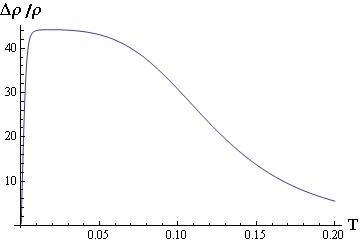} 
	 \caption{The plot depicts the magnetoresistance for a for heavily overdoped sample at lower temperature, 
	 which is to be contrasted to Fig. 1 of \cite{Hussey1996}.}
	 \label{fig:MagnetoPlots}
\end{center}
\end{figure}

The magnetoresistance is defined as follows:
\begin{align}
{\Delta \rho\over \rho}\equiv {\rho_{yy}(B)-\rho_{yy}(0)\over \rho_{yy}(0)} \;.
\end{align}
Unlike overdoped TBCO ($T_c\sim 30$ K), in optimally doped TBCO ($T_c \sim 80$ K) the weak magnetic field regime extends up to 60 T.
This has implications on the doping dependence of $\bb$.
The scaling dependence of the resistivity on magnetic field, via the scaling in equation (\ref{bcond}) is in qualitative agreement
with experimental results \cite{tyler}. Hence, magnetic fields,  which are in the linear regime at optimal doping are in fact
in the non-linear regime in the overdoped  region (optimal doping here is $E_b\to\infty$).

The magnetoresistance in heavily overdoped TBCO increases gradually with decreasing $T$, approaching a finite value at the lowest temperatures measured, around $30K$, in the low temperature and weak field regime
(being proportional to square of magnetic field); see Fig. 1 of \cite{Hussey1996}. This behavior is captured by our results, as depicted in Fig. \ref{fig:MagnetoPlots}. { We expect the strong dip as $T\to 0$ may be also visible if experiments at lower temperatures are performed. }

\subsection{Hall coefficient}

\begin{figure}[!ht]
\begin{center}
	 \includegraphics[width=0.65\textwidth]{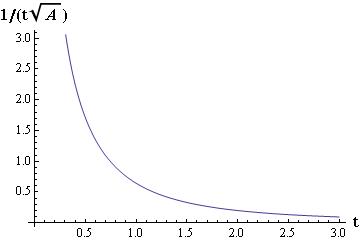}
	 \caption{Temperature dependence of the normalized Hall coefficient.
	 This corresponds to the function $\frac{1}{t \sqrt{A}}$ of our model. Compare this to the plot
	 of the quantity, ${R_H(T/T_*)-R_H(\infty)\over R^*_H}$, Fig. 2 of \cite{hwang}.
	 }
	 \label{fig:HallCoeffPlot}
\end{center}
\end{figure}

Attempts to identify a universal scaling behavior for the Hall coefficient in cuprate superconductors have not been very successful \cite{Ong1991}.
On the other hand, the inverse Hall angle depicts a universal behavior \cite{Ong1991}. It has been argued however, that the central anomaly of the Hall effect resides in direct measurements of the Hall coefficient \cite{Ong1997}.

To the best of our knowledge, there is only one report where a scaling behavior of the Hall coefficient
${R_H(T/T_*)-R_H(\infty)\over R^*_H}$, was argued to show a universal scaling behavior \cite{hwang}.
Here, $R_H(\infty)$ is the high temperature limit of $R_H$, $R^*_H$ rescales
the magnitude, and $T^*$ is a temperature scale. The scaling behavior is shown in Fig. 2 of \cite{hwang}. We compare this result to $\frac{1}{t \sqrt{A}}$ of our model,
which can be shown to be the Hall coefficient at a vanishingly small $B$ with only temperature scaling.
This is presented on Fig. \ref{fig:HallCoeffPlot}.

\subsection{K\"ohler rule}

K\"ohler's rule for metals states that $K=\rho^2{\Delta \rho\over \rho}$
should be independent of temperature. This was claimed to fail for YBCO and LSCO \cite{harris}. The authors of \cite{harris}
suggested, however, that a modified K\"ohler rule is valid and
$(\cot\Theta_H)^2 {\Delta \rho\over \rho}$ is approximately constant with temperature.

It has been argued that for LSCO  superconducting fluctuations play an important role in accounting for the difference between K\"ohler's rule and the modified K\"ohler rule \cite{kimura}.
While in principle our model can be shown to exhibit a superconducting transition by coupling to gauge and scalar fields, in the current setup our system does not include superconducting fluctuations.
Furthermore, at very low temperatures in the overdoped regime we do not expect that two such scales exist, as suggested by the very low temperature measurements of magnetoresistance.

Our data for the K\"ohler ratio and the modified K\"ohler ratio are in general temperature dependent, but not in the small temperature and large temperature limits.
Indeed, the facts that the resistivity and the inverse Hall angle are proportional at low temperatures and the modified K\"ohler ratio is constant implies that the K\"ohler ratio is also constant at very low temperatures.
{ Although this seems to be in contradiction with claims in the literature, we believe it should be valid at very low temperatures, in view of the proportionality of $\cot\Theta_H$ to $\rho$ \cite{mackenzie}.}

\section{Outlook}

A simple holographic system, namely the AdS-Schwarzschild black hole in light-cone coordinates, provides a solvable quantum critical model of magnetotransport with a wide range of properties.
The results obtained are in good agreement with those of strange metals, in particular the high-$T_c$ cuprates at very low temperatures with charge carrier concentration ranging from the optimal to the overdoped regime.
 An intriguing novel property emerging from our work is the scaling of the carrier doping dependence, hence, the model at $T = 0$  should be considered as a quantum critical line albeit with a Lifshitz scaling of z=2, which presents a radical departure from the paradigm of the isolated critical point.
This controls the linear resistivity in this regime as suggested in \cite{Hartnoll:2009ns}. Recent experimental results also point in this direction. This regime crosses over to a quadratic one, controlled by a standard CFT liquid. The crossover temperature diverges at optimal doping $E_b\to \infty$ explaining the high temperature reach of the linear resistivity regime.

Moreover, our findings provide several novel experimental and testable signatures for the low-temperature behavior of strange metals.
\begin{itemize}
\item The magneto-resistance vanishes abruptly near $T=0$.
\item  At sufficiently low temperatures, the transport data scale with a function $B/B_*$, where $B_*$ is doping dependent.
\item The K\"ohler rule and modified K\"ohler rule are both valid at low temperatures.
\end{itemize}
An extension to this work would be to clarify the underlying dynamics and how it matches the expected interactions of electrons in real materials.

The precise relation of the holographic model presented here with microscopic dynamics has yet to be clarified. Ideas in this direction have already been discussed \cite{sachdev} and connected to critical points and phases of the Hubbard model in \cite{Sachdev:2010uz}.
They are based on expectations of emergent strong non-abelian interactions at low energies and the ensuing holographic description. However, the non-standard holographic realization of the non-relativistic scaling symmetries remains a generic puzzle.
The emergence of superconductivity in this context is another important direction to be explored. For instance, using a probe scalar and gauge fields one would be able to study the onset of superconductivity and its dependence on quantum tuning parameters including charge carrier doping and magnetic field.

\addcontentsline{toc}{section}{Acknowledgments}
\section*{Acknowledgments}
We would like to thank T. Hu, M. Lippert, A. O'Bannon and D. Yamada for discussions.
This work has been partially supported by grants MEXT-CT-2006-039047, EURYI, FP7-REGPOT-2008-1-CreteHEPCosmo-228644,
 PERG07-GA-2010-268246 and the National Research Foundation, Singapore.

\appendix
\section*{Appendix}
\addcontentsline{toc}{section}{APPENDIX\label{app}}
\section{Detailed calculation of the conductivity}

\subsection{Conductivity calculation with $E^y_b$ and $E^z_b$}\label{sec:rotationInvariance}

In this section, we calculate the conductivity tensor in the absence of a
 magnetic field, and in the presence of arbitrary $E^y_b$ and $E^z_b$.
In this way we will also check that the system remains rotationally symmetric.
 The calculation is done in the probe approximation\footnote{This is
a good approximation strictly speaking when the number of flavors
remains fixed as the number of colors is large.} following \cite{Karch:2007pd}.
 We confirm that the light-cone electric field we turn on does not break rotational invariance.
Therefore, in the main part of the paper  $E_b$ should
be {taken} as $\sqrt{({E_b }^y)^2+({E_b }^z)^2}$.

For our purposes, we consider the ansatz
\begin{gather}
	A_+ = {E_b }^y y +{E_b }^z z + h_+  (r) \,, \qquad  A_- = b^2 {E_b}^y y +b^2 {E_b}^z z + h_-  (r)
	\,, \nonumber \\
	 A_y = b^2 E_b^y  x^- +  h_y (r) \,,  \qquad \quad A_z = b^2 E_b^z x^- +  h_z (r) \,,
\end{gather}
With these gauge fields, we have light-cone electric fields along $y, z$ directions.
For simplicity we set $2\pi \ell_s^2 = 1$ and ignore the contribution from the extra dimensions of $S^5$ as they are not relevant.

The DBI action is
\be
 S_{D_7} =- {\cal N} \int dr \sqrt{-det(g_{\mu\nu}+F_{\mu\nu})} =  - {\cal N} \int dr \sqrt{-detM}\;,
 \ee
  where $F_{\mu\nu}=\partial_{\mu}A_{\nu}-\partial_{\nu}A_{\mu}$ and
\begin{align}
	& detM =    g_{--}  \left( E_b^z h_y'  -E_b^y h_z' \right)^2+ G_{+-} g_{yy}  \left(h_y ^{' 2}+h_z^{' 2} \right)
	+\left((E_b^y)^2+(E_b^z)^2\right) g_{xx}  \left(g_{rr}  g_{--} +h_{-}^{' 2} \right)
	\nonumber\\
	& \qquad \qquad +g_{xx} ^2 \left(G_{+-} g_{rr} +g_{--}  h_{+}^{' 2} +h_{-}'  \left(-2 g_{+-}  h_{+}' +g_{++}  h_{-}' \right)\right)
	 \,,
\end{align}
with
\begin{align}
&G_{+-} =  -g_{+-} ^2+ g_{++}  g_{--}  \,,  \qquad
G_{+-y}=   \left( (E_{b}^y)^2+(E_{b}^z)^2 \right) g_{--} +G_{+-} g_{xx}  \;.
\end{align}
and the metric coefficients are defined in (\ref{AdSinlightcone}).

Varying the gauge fields, we obtain the constants of motion, (that are the currents)
\begin{align}
	&\langle J^+ \rangle  = \bar H \sp \bar H = - {\cal N} / \sqrt{-detM}
	g_{yy} ^2 \left(g_{--}  h_{+}' -g_{+-}  h_{-}' \right) \;, \nonumber \\	
	&\langle J^- \rangle =- \bar H \,
	 g_{yy}  \left(-g_{+-}  g_{yy}  h_{+}' +\left((E_b^y)^2+(E_b^z)^2+g_{++}  g_{yy} \right) h_{-}' \right)
	 \;,  \nonumber \\
	&\langle J^y \rangle  = \bar H  \, \left(G_{+-} g_{yy}  A_{y}' +(E_b^z) g_{--}  \left((E_b^z) A_{y}' -(E_b^y) A_{z}' \right)\right)  \;, \nonumber \\
	&\langle J^z \rangle  = \bar H  \, \left(G_{+-} g_{yy}  A_{z}' +(E_b^y) g_{--}  \left(-(E_b^z) A_{y}' +(E_b^y) A_{z}' \right)\right) \;, 	
\label{eqq}\end{align}

One solves the equations (\ref{eqq}) and substitute the solution back into the action to obtain the on-shell action
\begin{align}
	S_{D_7} &= - {\cal N}^2 \int dr~    g_{yy}
	\sqrt{g_{rr} }
	\sqrt{\frac{ G_{+-y}}{-   g_{yy}  {\cal N}^2- \bar U
	- \bar V }} \;,
\label{26}\end{align}
where
\begin{align}
	&\bar U  = \frac{(\langle J^y \rangle E_b^y +\langle J^z \rangle E_b^z )^2 g_{--}  + \left(\langle J^z \rangle^2+\langle J^y \rangle^2\right) G_{+-} g_{yy} }{G_{+-} G_{+-y} } \;, \\
	&\bar V  =\frac{\langle J^+ \rangle^2 \left( (E_b^y)^2+ (E_b^z)^2 \right)+\langle J^+ \rangle^2 g_{++}  g_{yy}  \langle J^- \rangle \{ 2 \langle J^+ \rangle g_{+-} +\langle J^- \rangle g_{--}  \} g_{yy} }{G_{+-y} g_{yy} }  \;.
\end{align}
and where  $\langle J^{\pm}\rangle,\langle J^{y,z}\rangle $ are constants and
\begin{align}
G_{+-y}(r)&= \left( (E_{b}^y)^2+(E_{b}^z)^2 \right) g_{--}(r) +G_{+-} g_{xx}(r)  \;.
\end{align}

As $r$ varies from the boundary of the geometry to the horizon, both the numerator and
 denominator in the square root in (\ref{26}) decrease and at some point change sign from positive to negative.
 Consistency for the solution implies that they should change sign at the same radial distance \cite{Karch:2007pd}.
We call $r_*$ the value of $r$ where this numerator changes the sign $G_{+-y}(r_*)=0$.
The functions $\bar U , \bar V $ have vanishing denominators, and
therefore we demand $\bar U(r_*)= \bar V(r_*) = 0$ at least as fast as $G_{+-y}$.
These conditions imply
\begin{align}
	\langle J^- \rangle = -\frac{g_{+-}  }{g_{--}  }
	\bigg\vert_{r=r_*}  \langle J^+ \rangle \,,  \qquad
	\langle J^z \rangle = \frac{- E_{b}^y  E_{b}^z  g_{--} }{(E_{b}^z)^2 g_{--} + G_{+-} g_{yy} }  \bigg\vert_{r=r_*}  \langle J^y \rangle \,. \label{z+currentrelation2}
\end{align}
Substituting this condition in the action and demanding that the denominator is zero at $r=r_*$,
we obtain the current along the $y$ direction. This equation gives
\begin{align}
\langle J^y \rangle^2 &=(E_b^y)^2 \frac{ \left( \langle J^+ \rangle ^2+   {\cal N}^2    g_{--}  g_{yy} ^2\right)}{g_{yy} ^2} \bigg\vert_{r=r_*} \;,
\label{27}\end{align}
where we used $G_{+-y}=0$ at $r=r_*$.
From equation (\ref{z+currentrelation2}), we obtain
\begin{align}
\langle J^z \rangle^2 &= (E_b^z)^2 \frac{ \left( \langle J^+ \rangle ^2+   {\cal N}^2    g_{--}  g_{yy} ^2\right)}{g_{yy} ^2} \bigg\vert_{r=r_*} \;.
\label{28}\end{align}

From the definition of the conductivity
\be
\langle J^i \rangle= \sigma_{ij} E_b^j \;,
\ee
and (\ref{27}) and (\ref{28}) we obtain
\be
\sigma^{yy} = \sigma^{zz}=\sqrt{\frac{ \left( \langle J^+ \rangle ^2+   {\cal N}^2    g_{--}  g_{yy} ^2\right)}{g_{yy} ^2} \bigg\vert_{r=r_*}}
\sp \sigma^{yz} = \sigma^{zy}=0 \;.
\label{30}\ee

Equation (\ref{30}) is the same as equation (\ref{conductivity}) in the main text after redefinitions.
It is clearly rotationally invariant with $(E_{b}^y)^2+(E_{b}^z)^2 \longrightarrow E_b^2$ .

\subsection{Hall conductivity calculation}\label{sec:HallCal}

In the presence of a magnetic field, we will  use the DBI probe technique developed in \cite{O'Bannon:2007in}.
The calculations are similar to the previous section, yet more involved.
Here we will present only the important steps in the calculation. We also take $2\pi \ell_s^2=1$ and
ignore the contribution from the extra dimensions of $S^5$ because it is not relevant.

To calculate the Hall conductivity, we choose the following gauge fields
\begin{eqnarray}
	A_+ = {E_b } y + h_+ (r) \,, ~~ A_- = 2 b^2 {E_b} y + h_- (r)
	\,, ~~ A_y = 2 b^2 E_b  x^+ +  h_y (r) \,, ~~ A_z = B_b y +  h_z (r) \,,
\end{eqnarray}
which are the same as equation (\ref{GaugeHall}). These gauge fields describe a light-cone  electric field along the $y$ direction
and a magnetic field perpendicular to  the $y-z$ plane.

The action is
$ S_{D_7} = - {\cal N} \int dr \sqrt{-detM}$, where
\begin{align}
	detM = & G_{+-y}g_{rr}(r)+g_{yy}(r)^2 h_+'(r)
		\left(g_{--}(r) h_+'(r)-2 g_{+-}(r) h_-'(r)\right) \nonumber\\ 
 		 &+g_{--}(r) \left(B_b h_+'(r)-E_b h_z'(r)\right)^2+G_{+-} g_{yy}(r) \left(h_y'(r)^2+h_z'(r)^2\right) \nonumber\\
 		 &+h_-'(r) \left(G_{+y} h_-'(r)+2 B_b g_{+-}(r) \left(-B_b h_+'(r)+E_b h_z'(r)\right)\right) \,.
\end{align}
and
\begin{gather}
G_{+-y}=  [B_b^2 + g_{yy}(r)^2]G_{+-}+E_b^2 g_{--}(r) g_{yy}(r) \;, \quad
G_{+-} =  -g_{+-}(r)^2+ g_{++}(r) g_{--}(r) \;,  \nonumber \\
G_{+y} = B_b^2 g_{++}(r)+E_b^2 g_{yy}(r)+g_{++}(r) g_{yy}(r)^2 \;.
\end{gather}
The constants of motion are (with simplified notation $h_+ = h_+(r)$)
\begin{align}
	&\langle J^+ \rangle  = \bar H \,
	 \left(-g_{+-} \left(B_b^2+g_{yy}^2\right) h_-'+g_{--} \left(\left(B_b^2+g_{yy}^2\right) h_+'-B_b E_b h_z'\right)\right) \;, \nonumber \\	
	&\langle J^- \rangle =- \bar H \,
	\left(G_{+y} h_-'-g_{+-} \left(\left(B_b^2+g_{yy}^2\right) h_+'-B_b E_b h_z'\right)\right)
	 \;,  \nonumber \\
	&\langle J^y \rangle  = \bar H  \, G_{+-}  g_{yy} h_y'   \;, \nonumber \\
	&\langle J^z \rangle  = \bar H  \, \left(B_b E_b g_{+-} h_-'+G_{+-} g_{yy} h_z'+E_b g_{--} \left(-B_b h_+'+E_b h_z'\right)\right) \;,  \\
	&\text{where} \quad \bar H = - \frac{ {\cal N} }{\sqrt{-detM}} \;.\nonumber
\label{eqqq}\end{align}
We solve  the equations (\ref{eqqq}) and substitute the solutions into the action, to obtain
\begin{align}
	S_{D_7} &= - {\cal N}^2 \int dr~
	\sqrt{\frac{ g_{rr} ~G_{+-y}}{-{\cal N}^2-\bar W(r)+ \bar U(r) - \bar V(r)}} 	
	  \;,
\end{align}
where
\begin{align}
	&\bar U(r) = \frac{-B_b^2 \langle J^+ \rangle^2 G_{+-}^2+E_b g_{--} \left(E_b (\langle J^z \rangle B_b+\langle J^+ \rangle E_b)^2 g_{--}+\langle J^+ \rangle (2 \langle J^z \rangle B_b+\langle J^+ \rangle E_b) G_{+-} g_{yy}\right)}{B_b^2 G_{+-} g_{--} \left(E_b^2 g_{--} g_{yy}+G_{+-} \left(B_b^2+g_{yy}^2\right)\right)}  \;, \nonumber \\
	&\bar V(r) = \frac{-(\langle J^+ \rangle g_{+-}+\langle J^- \rangle g_{--})^2}{g_{--} \left(-G_{+y} g_{--}+g_{+-}^2 \left(B_b^2+g_{yy}^2\right)\right)} \;, \nonumber \\
	&\bar W(r) = \frac{\left(\langle J^z \rangle^2+\langle J^y \rangle^2\right) B_b^2+2 \langle J^z \rangle \langle J^+ \rangle B_b  E_b+\langle J^+ \rangle^2 E_b^2}{B_b^2 G_{+-} g_{yy}} \;.
\end{align}

As before, we demand the square root factor to be real all the way from the horizon to the boundary.
The numerator of the action changes sign at $r_H < r=r_* < \infty$ and we solve it explicitly as
\be
\Big[ \left( g_{yy}(r)^2+B_b^2 \right) G_{+-}+E_b^2 g_{--}(r) g_{yy}(r)  \Big]_{r=r_*} = 0 \;,
\ee
which implies
\be
r_*^4 = \frac{1}{2} \left( r_H^4 -B_b^2 \ell^4 + \sqrt{ \left( r_H^2 + \ell^4 B_b^2 \right)^2 +4 E_b^2 b^2 r_H^4 \ell^4 }\right) \,.
\ee

For the on-shell action to be real, the denominator should also vanish at $r=r_*$.
It turns out that the functions $\bar U(r), \bar V(r)$ have also vanishing denominator.
Thus we demand $\bar U(r_*)= \bar V(r_*) = 0$ at least as fast as $G_{+-y}$.
Setting the numerators of $\bar V, \bar U$ to be zero at $r=r_*$, we obtain
\begin{align}
	&\langle J^- \rangle = -\frac{g_{+-}(r) }{g_{--}(r) }
	\bigg\vert_{r=r_*}  \langle J^+ \rangle \,,  \\
	&\langle J^z \rangle = - \frac{E_b^2 g_{--}(r)^2 + G_{+-} g_{--}(r) g_{yy}(r)}{B_b E_b g_{--}(r)^2} \bigg\vert_{r=r_*}  \langle J^+ \rangle \,. \label{z+currentrelation}
\end{align}

By plugging this condition to the denominator of the action,
we obtain the expression of the current along $y$ direction. In turn we use the Ohm's law to obtain
\begin{align}
	&\sigma^{yy} = \frac{g_{yy}(r)}{B_b^2+g_{yy}(r)^2} \sqrt{ \langle J^+ \rangle^2+ {\cal N}^2  g_{--}(r) \left(B_b^2+g_{yy}(r)^2\right)} \bigg\vert_{r=r_*}  \,.
\end{align}
This expression can be evaluated with explicit temperature dependence as
\begin{align}
	\sigma^{yy}  = &\frac{\ell}{{\cal G}_-} \left(64 \sqrt{2} \langle J^+ \rangle^2 {\cal G}_+ +\ell^2 {\cal N}^2 \pi ^4 T^4 b ^6 \sqrt{{\cal G}_+} {\cal G}_- ~\cos^6 \theta \right)^{1/2} \,,  \label{electricConductivitywithB2}  \\
	 &{\cal G}_+=  \ell^2 \sqrt{\left(B_b^2 + \pi^4 b^4 \ell^4  T^4 \right)^2 + 4 \pi^4 E_b^2 b^6 \ell^4 T^4 }  -B_b^2 \ell^2+ \pi^4 b^4 \ell^6 T^4 \;,\nonumber \\
	 &{\cal G}_-=  \ell^2 \sqrt{\left(B_b^2 + \pi^4 b^4 \ell^4  T^4 \right)^2 + 4 \pi^4 E_b^2 b^6 \ell^4 T^4 }  +B_b^2 \ell^2+ \pi^4 b^4 \ell^6 T^4 \;,\nonumber
\end{align}
where we used the identification $r_H = \ell^2 \pi  T b $ and these expressions are the same as the equations
(\ref{electricConductivitywithB}) and (\ref{bcond}) with appropriate identifications.
When the magnetic field vanishes, this expression reduces to the conductivity formula given in equation (\ref{conductivity}),
which provide a consistency check.
The Hall conductivity can be calculated from (\ref{z+currentrelation}) and (\ref{electricConductivitywithB2}) as
\begin{align}
\sigma^{yz}  = \frac{B_b \langle J^+ \rangle }{B_b^2+g_{yy}(r)^2}
 = \frac{2\ell^2  B_b \langle J^+ \rangle }{{\cal G}_-} \;,
\end{align}
which is identical to the equation (\ref{electricConductivitywithB}).
This Hall conductivity vanishes when the magnetic field vanishes.

\section{Study of the temperature dependence of the conductivity.\label{apb}}

In this appendix  we would  investigate  the temperature dependence of the conductivity
in various regimes in the presence of a magnetic field.
The two main regimes are the drag dominant regime (at lower temperatures) and also the ``pair creation"  regime at higher temperatures.

The basic starting formulae are
\begin{align}
\s_{yy}={\s_0\over \fm}\sqrt{\fp J^2+t^4\sqrt{\fp}\fm}\sp \s_{yz}=\bar\s_0{\bb\over \fm}\sp \cot\Theta _H={\s_{yy}\over \s_{yz}} \;,
\label{1} \\
{\cal F}_{\pm}=t^4\left[1\mp {\bb^2\over t^4}+\sqrt{\left(1+{\bb^2\over t^4}\right)^2+{1\over t^4}}\right]
\sp J^2={64\sqrt{2}\jp^2\over ({\cal N}b\cos^2\theta)^2(2b E_b)^3} \;,
\\
t={\pi \ell bT\over \sqrt{2b E_b}}\sp \bb={B_b\over 2bE_b}
\sp
\s_0={\cal N}b\cos^2\theta~\sqrt{2b E_b}\sp {\bar \s_0}={\jp\over bE_b} \;.
\end{align}

At small magnetic field,
\be
 {\cal F}_{\pm}=t^2{\cal A}(t)+\left({t^2\over \sqrt{t^4+1}}\mp 1\right)\bb^2+{\cal O}({\bb^4})\;,
 \ee
  and
the conductivities become
\begin{align}
\s_{yy}(0)=\s_0\sqrt{{J^2\over t^2{\cal A}}+{t^3\over \sqrt{\cal A}}}\sp \s_{yz}=0  \sp
{\cal A}(t)=t^2+\sqrt{1+t^4} \;.
\end{align}

\subsection{Drag dominated regime}

For drag dominated regime, we assume that the constant $J$ is large enough
 so that the $J$-independent result can be neglected.
We study the opposite case in the following subsection.
In this case
\be
\s_{yy}\simeq {J\s_0\over \fm}\sqrt{\fp}\sp \s_{yz}=\bar\s_0{\bb\over \fm} \;.
\label{1a}
\ee

Inverting  the conductivity tensor, we can derive the resistivity formula as
\begin{align}
\rho_{yy}={\s_0J \sqrt{\fp} \fm \over \s_0^2 J^2  \fp + \bar\s_0^2 \bb^2}\sp
\rho_{yz}={\bar\s_0 \bb \fm \over \s_0^2 J^2  \fp + \bar\s_0^2 \bb^2} \sp
\cot\Theta _H={\rho_{yy}\over \rho_{yz}}={\s_0 J\over \bar\s_{0} \bb}\sqrt{\fp} \;.
\end{align}
We will also calculate the rest of the related observables.
The magnetoresistance is defined as
\begin{align}
{\Delta \rho\over \rho}\equiv {\rho_{yy}(B)-\rho_{yy}(0)\over \rho_{yy}(0)} \;.
\end{align}
In the drag regime, it is equal to
\begin{align}
{\Delta \rho\over \rho}={ \sqrt{\fp} \fm  \over ( \s_0^2 J^2 \fp + \bar\s_0^2 \bb^2) t\sqrt{{\cal A}}  }-1
\simeq
\left( {\s_0^2 J^2 (3 + {t^2 \over \sqrt{t^4+1}} ) - 2 \bar\s_0^2  \over 2 \s_0^2 J^2 t^2 {\cal A}  }
\right) \bb^2+{\cal O}({\bb^4}) \;.
\label{MagnetoFormula}
\end{align}
Here we kept two terms in the denominator because they are at the same order in the drag limit.

In the weak field regime or linear field regime, which is defined as the regime with the
properties, $\frac{\Delta \rho}{\rho} \sim B_b^2$ and $\rho_{yz} \sim B_b$,
the magnetoresistance has the following behavior :
it diverges as $1/t^2$ as $t\to 0$ and vanishes as $1/t^4$ as $t\to \infty$.

The Hall resistance is
\begin{align}
R_H\equiv {\rho_{yz}\over B} ={\bar\s_0   \fm \over \s_0^2 J^2  \fp + \bar\s_0^2 \bb^2}
\simeq { \bar\s_0  \over \s_0^2 J^2  } +
\left( { \bar \s_0 (2 \s_0^2 J^2 -\bar \s_0^2)  \over  \s_0^4 J^4 t^2 {\cal A}  }
\right) \bb^2+{\cal O}({\bb^4}) \;.
\end{align}
In the  drag regime, we keep both terms in the denominator.
Overall, this is constant due to the first term. There are small corrections with temperature dependence.
In the small field regime it behaves as $t^{-2}$ as $t\to 0$ and $t^{-4}$ as $t\to \infty$.

We also define the K\"ohler ratio $K$, and the modified K\"ohler ratio $\tilde K$ as
\begin{align}
K=\rho_{yy}(0)^2{\rho_{yy}(B)-\rho_{yy}(0)\over \rho_{yy}(0)}=\rho_{yy}(0)(\rho_{yy}(B)-\rho_{yy}(0))
\sp
\tilde K=(\cot\Theta_H)^2{\Delta \rho\over \rho} \;.
\end{align}
In the drag regime we obtain for the K\"ohler ratio
\begin{align}
K\simeq { t \sqrt{{\cal A}} \over \s_0 J} {\Delta \rho\over \rho}+\cdots
\simeq
\left( {\s_0^2 J^2 (3 + {t^2 \over \sqrt{t^4+1}} ) - 2 \bar\s_0^2  \over 2 \s_0^4 J^4  }
\right) \bb^2+\cdots \;.
\end{align}
and for the modified K\"ohler ratio is
\begin{align}
\tilde K\simeq {\s_0^2J^2\over \bar \s_0^2 {\cal B}^2} \fp {\Delta \rho\over \rho} +\cdots
\simeq  {\s_0^2 J^2 (3 + {t^2 \over \sqrt{t^4+1}} ) - 2 \bar\s_0^2  \over 2 \bar \s_0^2}+\cdots \;.
\end{align}
For the two regimes, $t\ll 1$ and $t\gg 1$, $K$ and $\tilde K$ are both independent of $t$.

\subsubsection{Drag dominated regime I : $\bb \ll t^2$}

Using this condition we expand the square root as
\begin{align}
\sqrt{\left(1+{\bb^2\over t^4}\right)^2+{1\over t^4}}\simeq \sqrt{1+{1\over t^4}}\left[1+{\bb^2\over 1+t^4}+{\bb^4\over 2t^4(t^4+1)^2}+\cdots\right] \;.
\end{align}
To expand further we have to distinguish two cases

{\bf Ia}: $t\ll 1$. Then we have
\begin{align}
{\cal F}_{\pm}\simeq t^2\aa \mp \bb^2 +{\cal O}(\bb^4)\simeq t^2\cdots\mp \bb^2+\cdots \;.
\end{align}
Thus
\begin{align}
&\r_{yy}\simeq {1\over \s_0J } t\left(1+{\cal O}(t^2)+{3\over 2}{\bb^2\over t^2}+\cdots\right)
\sp \r_{yz}\simeq {\bar\s_0\bb \over \s_0^2 J^2 }+\cdots \;, \\
&\cot\Theta _H\simeq {\s_0 J\over \bar\s_{0} \bb}t+\cdots\sp {\Delta \rho\over \rho}\simeq {3\over 2}{\bb^2\over t^2}+\cdots \;.
\end{align}

{\bf Ib}: $t\gg 1$ and we obtain
\begin{align}
&\sqrt{\left(1+{\bb^2\over t^4}\right)^2+{1\over t^4}}\simeq 1+{1\over 2t^4}+{\bb^2\over t^4}+\cdots \;, \\
&\fp\simeq 2t^4\left[1+{1\over 4t^4}-{\bb^2\over 2t^8}+\cdots\right] \;, \qquad
\fm\simeq 2t^4\left[1+{1\over 4t^4}+{\bb^2\over t^4}+\cdots\right] \;.
\end{align}
Then
\begin{align}
&\r_{yy}\simeq {\sqrt{2}\over \s_0J } t^2\left[1+{1\over 8t^4}+{\bb^2\over t^4}+\cdots\right]
\sp \r_{yz}\simeq \r_{yz}\simeq {\bar\s_0\bb \over \s_0^2 J^2 } +\cdots \;, \\
&\cot\Theta _H\simeq {\sqrt{2}\s_0 J\over \bar\s_{0} \bb}t^2+\cdots\sp  {\Delta \rho\over \rho}\simeq {\bb^2\over t^4+{1\over 8}}+\cdots\;.
\end{align}

\subsubsection{Drag dominated regime II : $\bb\gg t^2$}

In this case the square root is expanded as
\begin{align}
\sqrt{\left(1+{\bb^2\over t^4}\right)^2+{1\over t^4}}\simeq \sqrt{{\bb^4\over t^8}+{1\over t^4}}\Big[1+\cdots\Big] \;.
\end{align}
Here we also distinguish two cases.

{\bf IIa}: $t\ll \bb~~~\to ~~~{\bb^4\over t^8}\gg{1\over t^4}$. Thus
\begin{align}
\fp\simeq 2t^4+{t^4\over 2\bb^2}+{t^8\over 2\bb^2}+\cdots \sp
\fm\simeq {2\bb^2}+2t^4+{t^4\over 2\bb^2}+{t^8\over 2\bb^2}+\cdots \;.
\end{align}
and
\begin{align}
\r_{yy}\simeq {2\sqrt{2}\s_0 J t^2 \over \bar\s_0^2 } \left(1+ {(1+9t^2) \bar \s_0^2 - 2 \s_0^2 J^2 t^4 \over 8 \bar\s_0^2 {\cal B}^2} +  \cdots \right) \sp \cot\Theta _H\simeq {\s_0 J\sqrt{1+4\bb^2}\over \bar\s_{0} \sqrt{2}\bb^2}~t^2    \;, \\
\r_{yz}\simeq {2\bb\over \bar\s_0 }+\cdots  \sp
{\Delta \rho\over \rho}\simeq  {2\sqrt{2}\s_0^2 J^2 t^2 \over \bar\s_0^2 t {\cal A} } \left(1+ {(1+9t^2) \bar \s_0^2 - 2 \s_0^2 J^2 t^4 \over 8 \bar\s_0^2 {\cal B}^2} +  \cdots \right) -1\;.
\end{align}

{\bf IIb}: $t\gg \bb~~~\to ~~~{\bb^4\over t^8}\ll{1\over t^4}$. This can only happen if $\bb\ll 1$ and this in turn implies that $t\ll 1$.
 We obtain
 \be
  {\cal F}_{\pm}\simeq t^2\left[1\mp{\bb^2\over t^2}+\cdots\right] \;,
  \ee
   and we have
\begin{align}
&\r_{yy}\simeq {1\over \s_0J }\left[t+{3\over 2}{\bb^2\over t}+\cdots\right]
\sp \r_{yz}\simeq  {\bar\s_0\bb \over \s_0^2 J^2 }+\cdots \;, \\
&\cot\Theta _H\simeq {\s_0 J\over \bar\s_{0} \bb}t+\cdots\sp {\Delta \rho\over \rho}\simeq {3\over 2}{\bb^2\over t^2}+\cdots \;.
\end{align}
Note that case IIb has the same asymptotics as case Ia.

\subsection{Pair creation dominated regime}

In this case, the term proportional to ${\cal N}$ dominates compared to the drag term and
the conductivities simplify to
\begin{align}
&\s_{yy}={\s_0 ~t^2~\fp^{1\over 4}\over \fm^{1\over 2}}\sp \s_{yz}=\bar\s_0{\bb\over \fm}\sp \cot\Theta _H={\s_0 ~t^2\over \bar\s_0\bb }\fp^{1\over 4}\fm^{1\over 2} \;, \\
&{\cal F}_{\pm}=t^4\left[1\mp {\bb^2\over t^4}+\sqrt{\left(1+{\bb^2\over t^4}\right)^2+{1\over t^4}}\right] \;.
\end{align}

\subsubsection{Pair creation dominated regime I : $\bb\ll t^2$}

We have two different regimes to consider.

{\bf Ia}: $t\ll 1$.
\begin{align}
{\cal F}_{\pm}\simeq t^2+\cdots \sp
\s_{yy}\simeq {\s_0 } t^{3\over 2}+\cdots
\sp \s_{yz}\simeq{\bar\s_0\bb \over t^2}+\cdots\sp
\cot\Theta _H\simeq {\s_0 \over \bar\s_{0} \bb}t^{7\over 2}+\cdots\;.
\end{align}

{\bf Ib}: $t\gg 1$.
\begin{align}
{\cal F}_{\pm}\simeq 2t^4+\cdots \sp
\s_{yy}\simeq{\s_0 } t +\cdots
\sp \s_{yz}\simeq {\bar\s_0\bb \over t^4}+\cdots\sp
\cot\Theta _H\simeq {\s_0 \over \bar\s_{0} \bb}t^{5}+\cdots \;.
\end{align}

\subsubsection{Pair creation dominated regime II : $\bb\gg t^2$}

We have again two different regimes to consider.

{\bf IIa}: $t\ll \bb~~~\to ~~~{\bb^4\over t^8}\gg{1\over t^4}$.
\begin{align}
&\fp\simeq 2t^4+\cdots \sp
\fm\simeq {2\bb^2}+\cdots \\
&\s_{yy}\simeq {\s_0\over \bb } t^{3}+\cdots
\sp \s_{yz}\simeq{\bar\s_0 \over 2\bb}+\cdots\sp
\cot\Theta _H\simeq {2\s_0 \over \bar\s_{0} }t^{3}+\cdots \;.
\end{align}

{\bf IIb}: $t\gg \bb~~~\to ~~~{\bb^4\over t^8}\ll{1\over t^4}$.

This can only happen if $\bb\ll 1$ and this in turn implies that $t\ll 1$.
\begin{align}
{\cal F}_{\pm}\simeq t^2 \;.
\end{align}
This is again as in case Ia.

\subsubsection{Condition for the drag dominated regime}

The condition for the drag term to dominate over the pair-creation term in the conductivity reads from (\ref{1})
\begin{align}
{t^4\fm\over \sqrt{\fp}}\ll J^2\;.
\end{align}
We will examine this condition in the three distinct regimes.
For the region {\bf I:  $\bb\ll t^2$}, we have
\begin{align}
&{\bf Ia}: t\ll 1 \quad \text{with}\quad  t^5\ll J^2 \;, \\
&{\bf Ib}:  t\gg 1  \quad \text{with}\quad   \sqrt{2}t^6\ll J^2 \;.
\end{align}
For the region {\bf   II:  $\bb\gg t^2$}, we have
\begin{align}
&{\bf IIa}: t\ll \bb ~~\to ~~{\bb^4\over t^8}\gg{1\over t^4}  \quad \text{with}\quad \sqrt{2}\bb^2t^2\ll J^2   \;, \\
&{\bf IIb}:  t\gg \bb~~~\to ~~~{\bb^4\over t^8}\ll{1\over t^4}  \quad \rightarrow \quad \bb\ll 1 \quad \rightarrow
\quad t\ll 1 \;.
\end{align}
Therefore {\bf IIb} implies case Ia.

\end{document}